\documentclass[preprintnumbers,article,amsmath,amssymb,floatfix,10pt,prd,twocolumn,
superscriptaddress,nofootinbib]{revtex4-2}
\usepackage{bm}
\usepackage{amsfonts}
\usepackage{latexsym}
\usepackage[latin1]{inputenc}
\usepackage{graphicx}
\usepackage{amsmath}
\usepackage[version=3]{mhchem}
\usepackage{palatino}
\usepackage{mathpazo}
\usepackage{textcomp}
\usepackage{float}
\usepackage{booktabs}
\usepackage{dcolumn}
\usepackage{ragged2e}
\usepackage{hyperref}
\hypersetup{colorlinks,citecolor=blue}
\hypersetup{colorlinks=true,linkcolor=red,filecolor=magenta,    urlcolor=blue}
\usepackage{amsmath}
\usepackage{xcolor}
\usepackage{orcidlink}
\usepackage{epsfig}
\usepackage[caption=false]{subfig}
\usepackage{commath}
\usepackage{caption}

\def\jnl@style{\it}
\def\aaref@jnl#1{{\jnl@style#1}}

\def\aaref@jnl#1{{\jnl@style#1}}

\def\aj{\aaref@jnl{AJ}}                   
\def\apj{\aaref@jnl{ApJ}}                 
\def\apjl{\aaref@jnl{ApJ}}                
\def\apjs{\aaref@jnl{ApJS}}               
\def\apss{\aaref@jnl{Ap\&SS}}             
\def\aap{\aaref@jnl{A\&A}}                
\def\aapr{\aaref@jnl{A\&A~Rev.}}          
\def\aaps{\aaref@jnl{A\&AS}}              
\def\mnras{\aaref@jnl{Mon.~Not.~Roy.~Astron.~Soc.}}             
\def\prd{\aaref@jnl{Phys.~Rev.~D}}        
\def\prc{\aaref@jnl{Phys.~Rev.~C}}  
\def\prl{\aaref@jnl{Phys.~Rev.~Lett.}}    
\def\qjras{\aaref@jnl{QJRAS}}             
\def\skytel{\aaref@jnl{S\&T}}             
\def\ssr{\aaref@jnl{Space~Sci.~Rev.}}     
\def\zap{\aaref@jnl{ZAp}}                 
\def\nat{\aaref@jnl{Nature}}              
\def\aplett{\aaref@jnl{Astrophys.~Lett.}} 
\def\apspr{\aaref@jnl{Astrophys.~Space~Phys.~Res.}} 
\def\physrep{\aaref@jnl{Phys.~Rep.}}      
\def\physscr{\aaref@jnl{Phys.~Scr}}       
\def\commat{\aaref@jnl{Comm.~Math.~Phys.}}              
\def\science{\aaref@jnl{Science}}               
\def\cqg{\aaref@jnl{Classical Quant.~Grav.}}            
\def\jpcs{\aaref@jnl{JPCS}}                                     
\def\ijmpd{\aaref@jnl{Int.~J.~Mod.~Phys.~D}}                    
\def\grg{\aaref@jnl{Gen.~Relat.~Gravit.}}               
\def\rpp{\aaref@jnl{Rep.~Prog.~Phys.}}          
\def\npa{\aaref@jnl{Nucl.~Phys.~A}}        
\def\lrr{\aaref@jnl{Living Rev.~Rel.}}                   
\def\jcap{\aaref@jnl{J.~Cosmology Astropart.~Phys.}}    
\def\rmp{\aaref@jnl{Rev.~Mod.~Phys.}}   
\def\epjc{\aaref@jnl{Eur.~Phys.~J.~C}} 
\def\plb{\aaref@jnl{~Phy.~Lett.~B}} 
\def\mpla{\aaref@jnl{Mod.~Phy.~Lett.~A}} 
\def\arxiv{\aaref@jnl{arxiv.org}}

\allowdisplaybreaks[1]

\begin{document}

\color{black}       
\title{Big Bang Nucleosynthesis constraints on $f(T, \mathcal{T})$ gravity}

\author{Sai Swagat Mishra\orcidlink{0000-0003-0580-0798}}
\email{saiswagat009@gmail.com}
\affiliation{Department of Mathematics, Birla Institute of Technology and
Science-Pilani,\\ Hyderabad Campus, Hyderabad-500078, India.}

\author{Ameya Kolhatkar\orcidlink{0000-0002-9553-1220}}
\email{kolhatkarameya1996@gmail.com}
\affiliation{Department of Mathematics, Birla Institute of Technology and
Science-Pilani,\\ Hyderabad Campus, Hyderabad-500078, India.}

\author{P.K. Sahoo\orcidlink{0000-0003-2130-8832}}
\email{pksahoo@hyderabad.bits-pilani.ac.in}
\affiliation{Department of Mathematics, Birla Institute of Technology and
Science-Pilani,\\ Hyderabad Campus, Hyderabad-500078, India.}
%
\date{\today}
\begin{abstract}
\noindent Big Bang Nucleosynthesis provides us with an observational insight into the very early Universe. Since this mechanism of light element synthesis comes out of the standard model of particle cosmology which follows directly from General Relativity, it is expected that any modifications to GR will result in deviations in the predicted observable parameters which are mainly, the neutron-to-proton ratio and the baryon-to-photon ratio. We use the measured neutron-to-proton ratio and compare the theoretically obtained expressions to constrain two models in the framework of $ f(T,\mathcal{T}) $ gravity. The theoretically constrained models are then tested against observational data from the Hubble dataset and the $ \Lambda $CDM model to explain the accelerated expansion of the Universe.\\

\textbf{Keywords:} Big Bang Nucleosynthesis, Neutron abundance, Early cosmology, $f(T,\mathcal{T})$ gravity.
\end{abstract}

\maketitle

\date{\today}

\section{Introduction}
The observational discovery of the accelerated expansion of the Universe \cite{super1,super2,cmb1,cmb2,cmb3,cmb4,cmb5,cmb6,bao,weaklens} is enough to make modifications to conventional General Relativity (GR) and correspondingly the concordance model of cosmology. Various models have been proposed and explored in order to account for the large scale behaviour that seems to be missing from GR. They include -- scalar-tensor theories, direct extensions to GR like the $ f(R) $ class of gravities, teleparallel equivalents to GR and their extensions (the $ f(T) $ and $ f(Q) $ classes of gravities), massive gravities, theories with non-minimally coupled geometry and matter, and so on. All these efforts have been made in order to address the issues of the large scale structure, horizon problem, fine-tuning problem, matter-antimatter asymmetry, and the $ H_0 $ and $ \sigma_8 $ tensions which are of a $ 5\sigma $ significance. From a plethora of available candidates, we now turn to a specific class of theories called $ f(T) $ gravity \cite{Linder/2010,Salako/2013,Bamba/2012,Hamani/2011,Tamanini/2012,Ferraro/2011,Boehmer/2012}, where $ T $ is the torsion scalar. It is to be noted that when $ f(T)=T $, the theory is equivalent to GR up to a boundary term and hence it is famously dubbed as the Teleparallel Equivalent of General Relativity \cite{TEGR}. The $ f(T) $ framework can be further extended to the $ f(T,\mathcal{T}) $ framework which was introduced in \cite{harko} where $ \mathcal{T} $ corresponds to the trace of the energy-momentum tensor. By coupling this torsion to the trace of the energy-momentum tensor, one can get an interesting behaviour that explains the different epochs of the evolution of the Universe \cite{fTTdavood}. For more on work in this gravity, check out \cite{fTT1,fTT2,fTT3,fTT4,fTT5,fTT6}.  

The phase of Big Bang Nucleosynthesis (BBN), which occurred within a few minutes after the big bang, was the one in which the light nuclear elements (D, \ce{^{3}He}, \ce{^{4}He}, \ce{^{7}Li}) were created. The relative abundances of light elements can be calculated theoretically and then matched with the observed values. As far as the standard model of cosmology is concerned, the two match quite well. When we turn to modified gravities, however, geometrical effects due to non-linear and/or non-minimally coupled terms induce different early Universe scenarios than those found in the standard model. Observations can thus be used to extract constraints on various model parameters of the said models. BBN provides us with direct observational evidence in the form of neutron-to-proton ratio and the baryon-to-photon ratio that can constrain cosmological models (refer \cite{bbn1, Fields/2020,Barrow/2021,Anagnostopoulos/2023,Kang/2009,
Capozziello/2017} for a detailed report on BBN). This is necessary because any predictive late-time model or theory must satisfy these early time constraints thrust upon it.

The manuscript is organized as follows -- after an introduction to the formalism of $ f(T,\mathcal{T}) $ gravity in section II, we move to the basics of constraints obtained from BBN in section III. Section IV explores these constraints in the context of $ f(T,\mathcal{T}) $ gravity. Section V aims to explain the late time accelerated era by a thorough comparison of our constrained model with the Cosmic Chronometer dataset and $\Lambda$CDM model. We finally end the manuscript with the conclusion in section VI.
\section{Formalism of $f(T,\mathcal{T})$ gravity }
In this section, we will discuss the fundamental equations required in $f(T,\mathcal{T})$ gravity theory. To obtain a torsion-based curvature, the required connection is called Weitzenb\"ock \cite{Weitzenbock} connection, which is defined as
\begin{equation}
\label{1}
\overset{w}{\Gamma}_{\nu \mu}^\lambda \equiv e_{A}^\lambda\partial_{\mu}e_{\nu}^A=-e_{\mu}^A \partial_{\nu} e_{A}^\lambda
\end{equation}
 Here $e_{A}^\lambda$ and $e_{\nu}^A$ are tetrads. Basically, this connection leads to zero curvature instead of zero torsion. The metric tensor related to these tetrads is \begin{equation}
 \label{2}
 g_{\mu \nu}(x)=\eta_{AB}e_{\mu}^A(x)e_{\nu}^B(x)
 \end{equation}
 here the Minkowski metric $\eta_{AB}=diag(1,-1,-1,-1)$.

From the above connection, geometrical objects like torsion tensor, contorsion tensor, and superpotential tensor can be obtained. The torsion tensor is  defined as,
\begin{equation}
\label{3}
    T_{\mu \nu}^\lambda=\overset{w}{\Gamma}_{\nu \mu}^\lambda-\overset{w}{\Gamma}_{\mu \nu}^\lambda=e_{A}^\lambda(\partial_{\mu}e_{\nu}^A-\partial_{\nu}e_{\mu}^A).
\end{equation}
From the torsion tensor,  contorsion tensor can be obtained as  ${K^{\mu \nu}}_{\rho} \equiv -\frac{1}{2}({T^{\mu \nu}}_{\rho}-{T^{\nu \mu}}_{\rho}-{T_{\rho}}^{\mu \nu})$. Using the two geometrical objects torsion and contorsion, another tensor can be obtained which is called superpotential tensor ${S_{\rho}}^{\mu \nu}$. It is defined as
\begin{equation}
\label{4}
    {S_{\rho}}^{\mu \nu} \equiv \frac{1}{2}({K^{\mu \nu}}_{\rho}+\delta_\rho^\mu {T^{\alpha \nu}}_{\alpha}-\delta_\rho^\nu {T^{\alpha \mu}}_{\alpha}).
\end{equation}
One can construct the torsion scalar $T$ using the Torsion tensor \eqref{3} and superpotential tensor \eqref{4} as follows:
\begin{equation}
\label{5}
    T\equiv  {S_{\rho}}^{\mu \nu} T_{\mu \nu}^\rho
      = \frac{1}{4}{T^{\rho \mu \nu}T_{\rho \mu \nu}}+ \frac{1}{2}{T^{\rho \mu \nu}T_{\nu \mu \rho}}-{T_{\rho \mu}}^\rho {T^{\nu \mu}}_\nu.
\end{equation}
In the modified version of teleparallel gravity, one can extend $f(T)$ to a general function of the torsion scalar $T$ and trace of energy-momentum tensor $\mathcal{T}$. The modified gravitational action for $f(T,\mathcal{T})$ gravity can be defined as,
\begin{equation}
\label{6}
     S=\frac{1}{16\pi G}{\int{{d^4}xe[T+f(T,\mathcal{T})]+\int{{d^4}xe\mathcal{L}_{m}}}}
\end{equation}
where $e=det(e_\mu^A)=\sqrt{-g}$, $G$ denotes the Newton's constant and $\mathcal{L}_m$ is the matter Lagrangian.\\
We can get the field equations, by varying the action \eqref{6} with respect to the tetrads, as
\begin{multline}
 \label{7}
   (1+f_T)\left[e^{-1}\partial_\mu(e\,{e_A}^\alpha S_\alpha^{\rho\mu})-e_A^\alpha T^{\mu}_{\nu \alpha} S_{\mu}^{\nu \rho} \right]+\\
   \left(f_{TT}\, \partial_{\mu} T+ f_{T \mathcal{T}}\,\partial_{\mu} \mathcal{T}\right)e\,e_{A}^{\alpha} S_{\alpha}^{\rho \mu}+e_{A}^\rho\left(\frac{f+T}{4}\right)\\
-\frac{f_{\mathcal{T}}}{2} \left(e^{\alpha}_A \stackrel{em}{T}_\alpha^{\rho} +p\, e_A^{\rho} \right)= 4\pi G\, e^{\alpha}_A \stackrel{em}{T}_\alpha^{\rho}
\end{multline}
 where ${\stackrel{em}{T}_\alpha}^\rho$ denotes the energy-momentum tensor, $f_T={\partial f}/{\partial T}$,  $f_{TT}={\partial^2 {f}}/{\partial T^2}$ and $f_{T\mathcal{T}}={\partial^2{f}}/{\partial T \partial \mathcal{T}}$.  In the context of perfect fluid description, the energy-momentum tensor has the diagonal form,
 \begin{equation}
 \label{8}
 {\stackrel{em}{T}_\alpha}^\rho=diag (\rho,\, -p,\,-p,\,-p)
 \end{equation}
 where $\rho$ and $p$ are energy density and the thermodynamic pressure respectively. Here  $\mathcal{T} = \rho-3p$.\\
Further discussion of the geometry of the universe will be proceeded by assuming a spatially flat Friedmann-Lemaitre-Robertson-Walker (FLRW) metric,
   \begin{equation}
   \label{9}
       ds^2=dt^2-a^2(t)\delta_{ij}dx^i dx^j,
   \end{equation}
   where a(t) is the scale factor as a function of time. For this metric, $T=-6H^2$ and the vierbein field is considered as the diagonal form, $e_\mu^A=diag(1,a,a,a)$ for the above metric.

Using the above metric \eqref{9} in the field equation \eqref{7}, we obtain the modified Friedmann equations:
\begin{equation}\
\label{10}
  {H^2=\frac{8\pi G}{3}{\rho}-\frac{1}{6}(f+12H^2f_T)+f_\mathcal{T}(\frac{\rho+p}{3})},
 \end{equation}
 
  \begin{multline}\
  \label{11}
  {\dot{H}=-4\pi G(\rho+p)}-\dot{H}(f_T-12H^2f_{TT})-\\H(\dot{\rho}-3{\dot{p}})f_{T\mathcal{T}}-f_\mathcal{T}(\frac{\rho+p}{2}).
  \end{multline}
  
  The conservation equation for ordinary matter can be written as
  \begin{equation}
      \label{12}
      \dot{\rho}+3H(\rho+p)=0
  \end{equation}
  and the equation of state reads as $p=\omega \rho$, where $\omega$ is the equation of state parameter. By using the Hubble parameter $H=\dot{a}/a$ and the above equation of state in the conservation equation \eqref{12} one can obtain the energy density in terms of scale factors as
  \begin{equation}
\label{13}
\rho=\frac{\rho_{0}}{a^{3(1+\omega)}}={\rho_{0}}(1+z)^{3(1+\omega)},
\end{equation}
where $\rho_{0}$ is the present density and the redshift $z=\frac{1}{a} -1$.\\
The Friedmann equations \eqref{10} and \eqref{11}, including the dark component can be written as 
\begin{equation}
\label{14}
    H^2=\frac{8 \pi G}{3} (\rho + \rho_{DE})
\end{equation}
\begin{equation}
\label{15}
    \dot{H}=-4 \pi G(\rho + \rho_{DE}+ p + p_{DE})
\end{equation}
where 
\begin{equation}
\label{16}
\rho_{DE}=\frac{1}{8\pi G}\left[-6 H^2 f_T -\frac{f}{2}+f_{\mathcal{T}}(\rho + p)\right]
\end{equation}
\begin{multline}
\label{17}
-4 \pi G p_{DE}= 4\pi G\rho_{DE}-\frac{f_{\mathcal{T}}}{2} (\rho + p)-\\H(\dot{\rho}-3\dot{p})f_{T \mathcal{T}}+12 H^2 \dot{H} f_{TT}-\dot{H}f_T
\end{multline}
\section{BBN CONSTRAINTS}
In the following section, we review the formalism of BBN within the $ f(T,\mathcal{T}) $ cosmology and hope to extract constraints on various model parameters. Since the neutron-to-proton ratio is affected due to expansion, it is important to study the effects of expansion on BBN. It is to be noted that the BBN event occurred during the radiation-dominated era. For a temperature $ T $ and the effective number of degrees of freedom given by $ g_* $, one can arrive at the energy density of the relativistic particles given by
\begin{equation}\label{18}
    \rho_r=\frac{\pi^2}{30}g_*T^4
\end{equation}
This $ \rho_r $ constitutes the energy density of radiation due to both relativistic particles and massless radiation. $ g_*\sim 10 $. Equation \eqref{14} for GR can be written as
\begin{equation}\label{19}
    H^2=\frac{1}{3M_p^2}\rho
\end{equation}
Here note that $ M_p=\frac{1}{\sqrt{8\pi G}}=1.22\times 10^{19}\;GeV $ is the reduced Planck mass\footnote{$M_{pl}=\sqrt{8\pi} M_p$ is the more conventionally used Planck mass} and $ \rho=\rho_m+\rho_r $. Since BBN takes place during the radiation dominated era, we can write it as
\begin{equation}\label{20}
    H^2\approx\frac{1}{3M_p^2}\rho_r\equiv H_{GR}^2
\end{equation}
In order to differentiate the Hubble rate in conventional GR with the one obtained in modified gravity, we label the prior with $ 
H_{GR} $ while the latter is denoted by just $ H $. These distinctions in notation shall be useful in finding out the deviations in the Hubble rate ($\Delta H$) which in turn can be used to calculate the deviations on the freeze-out temperature ($ 
\Delta T_f $) in order to put a constraint on the models. Thus \eqref{14} can be used along with \eqref{20} to get
\begin{equation}\label{21}
    H=H_{GR}\sqrt{1+\frac{\rho_{DE}}{\rho_r}}
\end{equation}
The quantity in the square-root can be expanded up to first order due to the fact that during the radiation dominated era, $ 
\rho_{DE}<<\rho_r $ so that
\begin{equation}\label{22}
    \Delta H=H-H_{GR}\approx\frac{\rho_{DE}}{\rho_r}\frac{H_{GR}}{2}
\end{equation}
We can now use the relation \eqref{18} to get the expression of the Hubble rate as a function of temperature.
\begin{equation}\label{23}
    H(T)=\sqrt{\frac{\pi^2 g_*}{90}}\frac{T^2}{M_p}
\end{equation}
A simple argument can be used to extract a relationship between temperature and time. In the radiation era, $ a(t)\sim t^{1/2} $ and consequently, $ H(t)\sim\frac{1}{2t} $ so that
\begin{equation}\label{24}
    \frac{1}{t}\sim\sqrt{\frac{2\pi^2 g_*}{45}}\frac{T^2}{M_p}
\end{equation}
A quantity that holds prime importance in the study of BBN is the freeze-out temperature $ T_f $. Neutrons and protons convert into one another via the weak interactions. If the ambient temperature is greater than $ 1\;MeV $, which is high compared to the expansion rate, these reactions observe equilibrium. But as the temperature drops below $ 1\;MeV $, the neutron-to-proton ratio ``freezes out" at about $ 1/6 $ only to slowly decrease by free neutron decay. Thus neutron abundance is calculated by knowing the neutron to proton $ \lambda_{pn}(T) $ and proton to neutron $ \lambda_{np}(T) $ conversion rates. Neutrons decay into protons (with electrons($ e^+ $), neutrinos($ \nu_e $) and anti-neutrinos($ \Bar{\nu}_e $) as by products ) through three different reactions $ n\longrightarrow p+e^-+\Bar{\nu}_e $, $ n+\nu_e\longrightarrow p+e^- $ and $ n+e^+\longrightarrow p+\Bar{\nu}_e $. So then 
\begin{multline}\label{25}
    \lambda_{pn}(T)=\lambda_{(n\longrightarrow p+e^-+\Bar{\nu}_e)}+\lambda_{(n+\nu_e\longrightarrow p+e^-)}+\\
    \lambda_{(n+e^+\longrightarrow p+\Bar{\nu}_e)}
\end{multline}
and its inverse $ \lambda_{np}(T) $ can be used to find the total conversion rate $ \lambda_{tot}(T)=\lambda_{pn}(T)+\lambda_{np}(T) $ to yield
\begin{equation}\label{26}
    \lambda_{tot}(T)=4AT^3(4!T^2+2\times 3!\mathcal{Q}T+2!\mathcal{Q}^2)
\end{equation}
here $\mathcal{Q}=m_n-m_p=1.29\times10^{-3} GeV$ and $ A=1.02\times 10^{-11} GeV^{-4} $. The freeze-out temperature corresponds to the following Hubble rate $ H(T_f)=\lambda_{tot}(T_f)\approx c_qT_f^5 $. Here $ c_q=4A4!\approx9.8\times 10^{-10}\;GeV^{-4} $. Using \eqref{23} at $ T=T_f $, 
\begin{equation}\label{27}
    T_f=\bigg(\frac{\pi^2 g_*}{90M_p^2c_q^2}\bigg)^{1/6}\approx 0.6\;MeV
\end{equation}
Following $ H(T_f)=c_qT_f^5 $, we have $ \Delta H=5c_qT_f^4\Delta T_f $. Inserting $ 
\Delta H $ from \eqref{22},
\begin{equation}\label{28}
    \frac{\Delta T_f}{T_f}\approx\frac{\rho_{DE}}{\rho_r}\frac{H_{GR}}{10c_qT_f^5}
\end{equation}
Incorporating the observational bound, we have the final constraint expression in terms of the freeze-out temperature
\begin{equation}\label{29}
    \bigg|\frac{\Delta T_f}{T_f}\bigg| < 4.7\times 10^{-4}
\end{equation}

\section{BBN constraints on $ f(T,\mathcal{T})$ gravity}
\subsection{Model 1: $  f(T,\mathcal{T})= \beta_1 T+  \beta_2 \mathcal{T}$}
Let us consider the Lagrangian form,
\begin{equation}
\label{30}
  f(T,\mathcal{T})=\beta_1 T+\beta_2 \mathcal{T}
\end {equation}
Using the functional form \eqref{30} in the first motion equation \eqref{10} one can obtain the model parameter $\beta_1$ corresponding to the present time as,
\begin{equation}\label{31}
    \beta_1=\frac{1}{3}\left(\Omega_{m_0}+\Omega_{r_0}-1+\frac{\beta_2 M_p^2}{2}\left(3\Omega_{m_0}+8\Omega_{r_0}\right)\right)
\end{equation}
where $\Omega_{m_0}$, $\Omega_{r_0}$ are the present density parameters for matter, relativistic particles respectively. Now substituting the model \eqref{30} in \eqref{16} gives us
\begin{equation}\label{32}
    \rho_{DE}= M_p^2\left(-3 \beta_1 H^2 +\frac{\beta_2}{2}(\rho+5p)\right)
\end{equation}
Since in the radiation dominated phase, $p_r = \rho_r/3$ (From equation of state $p=\omega \rho$) and $\rho_m << \rho_r$ we can rewrite the above equation as,
\begin{equation}\label{33}
    \rho_{DE}= M_p^2\left(-3 \beta_1 H^2 +\frac{4 \beta_2 \rho_r}{3}\right)
\end{equation}
Inserting the above energy density along with the density for relativistic particles \eqref{18} and \eqref{20} in \eqref{28} we obtain,
\begin{equation}\label{34}
    \frac{\Delta T_f}{T_f}=\frac{\pi}{90\sqrt{10}}\left(\frac{  \left(4 \beta _2 M_p^2-3 \beta _1\right) \sqrt{g_{*}}}{ M_p c_q T_f^3}\right)
\end{equation}
\begin{figure}[H]
    \centering
    \includegraphics[scale=0.95]{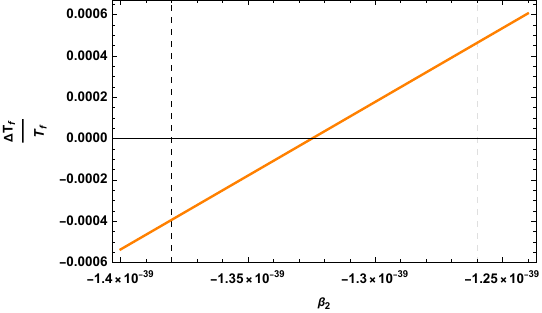}
    \caption{The model parameter $\beta_2$-dependence of $|{\Delta T_f}/T_f|$ for $ f(T,\mathcal{T})=\beta_1 T +\beta_2 \mathcal{T}$ .}
     \label{fig 1}
\end{figure}

The linear model passes the BBN constraints \eqref{29} for the values of model parameter $\beta_2 \in [-1.38\times 10^{-39},-1.26\times 10^{-39}]$. This range is represented by the dashed lines in Fig.\ref{fig 1}. For the calculation, we have incorporated the fixed parameters as $g_*\sim10$, $T_f=0.0006 \ GeV$, $c_q=9.8\times 10^{-10}\ GeV^{-4} $ and $M_p=1.22\times10^{19} \ GeV$. From \eqref{31} the range of the other model parameter can be obtained as $\beta_1 \in [-0.2640,-0.2615]$. Here the present values of the density parameters are taken from the observation \cite{{Aghanim/2020},
{Paulo/2020}} as $\Omega_{m_0}\sim0.3$ and $\Omega_{r_0}\sim0.00005$. Using \eqref{32} in the first motion equation we obtain,
\begin{multline}\label{35}
    H^2(z)=\frac{H_0^2}{1+3 \beta_1}\bigg(\Omega_{m_0}(1+z)^3(1+\frac{3\beta_2 M_p^2}{2})+ \\  \Omega_{r_0}(1+z)^4(1+4\beta_2 M_p^2)\bigg)
\end{multline}
here the relations $\Omega_{m}=\Omega_{m_0}/{a^3}$ and $\Omega_{r}=\Omega_{r_0}/{a^4}$  are used.

\subsection{Model 2: $  f(T,\mathcal{T})=T+\alpha T^2 +\gamma \mathcal{T}$}
Let us consider the Lagrangian form,
\begin{equation}
\label{36}
  f(T,\mathcal{T})=T+\alpha T^2 +\gamma \mathcal{T}
\end {equation}
Using the functional form in the first Friedmann equation \eqref{10} one can obtain the model parameter $\alpha$ corresponding to the present time as
\begin{equation}
    \label{37}
    \alpha=\frac{1}{18 H_0^2} \left[2- \Omega_{m_0} b_{m_0} - \Omega_{r_0} b_{r_0}\right]
\end{equation}
where $\Omega_{m_0}$, $\Omega_{r_0}$ are the present density parameters for matter, relativistic particles respectively, and $b_{m_0}=1+\frac{\gamma M_p^2}{2}$, $b_{r_0}=1+\frac{4 \gamma M_p^2}{3}$.
Substituting \eqref{36} along with \eqref{37} in the effective density \eqref{16} we get,

\begin{multline}\label{38}
    \rho_{DE}= M_p^2\bigg[ -3 H^2+ 
    \frac{3 H^4}{ H_0^2} \left(2-\Omega_{m_0} b_{m_0} - \Omega_{r_0} b_{r_0}\right) \\
     + \frac{\gamma}{2} (\rho + 5 p) \bigg]   
\end{multline}

Since in the radiation dominated phase, $p_r = \rho_r/3$ (From equation of state $p=\omega \rho$) and $\rho_m << \rho_r$ we can rewrite the above equation as,
\begin{multline}\label{39}
    \rho_{DE}= M_p^2\bigg[ -3 H^2+ 
    \frac{3 H^4}{ H_0^2} \left(2-\Omega_{m_0} b_{m_0} - \Omega_{r_0} b_{r_0}\right) \\
     + \frac{4 \gamma \rho_r}{3}  \bigg]   
\end{multline}
Substituting the above equation along with \eqref{18} and \eqref{20} in \eqref{28} we get,
\begin{multline}\label{40}
\frac{\Delta T_f}{T_f}=\frac{M_p}{10 \sqrt{3} c_q T_f^5} \bigg( -\frac{1}{M_p^2} + \frac{4 \gamma}{3} +\\
\frac{\pi T_f^2}{3 \sqrt{3} H_0^2 M_p^4}\left(2-\Omega_{m_0} b_{m_0} - \Omega_{r_0} b_{r_0}\right) \bigg)
\end{multline}
\\
\begin{figure}[H]
    \centering
    \includegraphics[scale=0.95]{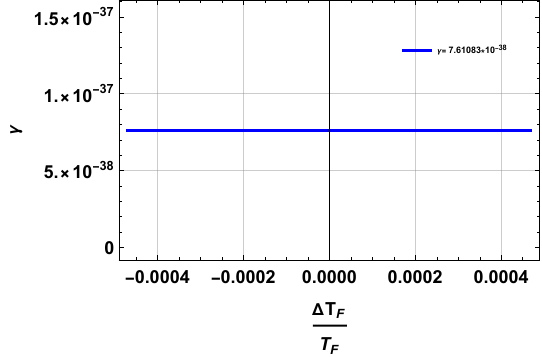}
    \caption{The model parameter $\gamma$-dependence of $|{\Delta T_f}/T_f|$ for $ f(T,\mathcal{T})=T+\alpha T^2 +\gamma \mathcal{T}$ .}
     \label{fig 3}
\end{figure}
The second model passes the BBN constraints \eqref{29} for the model parameter $\gamma=7.61083\times 10^{-38}$. In Fig.\ref{fig 3} we observe that the behavior of $\gamma$ is constant in the BBN region. The other model parameter $\alpha$ can be obtained from \eqref{37} by substituting the present observed values of the density parameters and the Hubble parameter $H_0 = 1.47\times 10^{-42\;}GeV$ ($\sim 67.2$ km $s^{-1}$ Mp$c^{-1}$)  as $\alpha=4.82\times 10^{82\;} GeV^{-2}$. The remaining fixed parameters are same as the first model.\\
Now, inserting the Lagrangian form \eqref{36} in the first motion equation \eqref{10} we obtain,
\begin{widetext}
\begin{equation}\label{41}
H^2(z)=H_0^2
  \left(\frac{ -1+\sqrt{1+  (1+z)^3 \left(\Omega_{m_0} b_{m_0} + \Omega_{r_0} b_{r_0}-2\right) \left( \Omega_{m_0} b_{m_0} + \Omega_{r_0} b_{r_0}(1+z)\right)}}{  \Omega_{m_0} b_{m_0} + \Omega_{r_0} b_{r_0} -2}\right) 
\end{equation}
\end{widetext}

\section{\textbf{Data Analysis and Results}}
Though our models have passed the BBN constraints for certain values of model parameters, its credibility concerning the evolution of the Universe is still unknown. So, in this section, we compare our models with the Cosmic Chronometer (CC) dataset and the standard $\Lambda$CDM model, using the constrained model parameters from the BBN epoch following the method in \cite{fTBBN}.
\subsection{\textbf{Cosmic Chronometer (CC) dataset}}
In order to account for the cosmic acceleration, various observational probes are used. To ensure that the probes quantifying cosmic evolution do not take into account effects of their own evolution, probes like standard candles (SNeIa) or standard rulers (Baryon Acoustic Oscillations) are widely used. We have employed the probes based on what are called ``Cosmic Chronometers" (first discussed in \cite{CCJimenez}) where relative ages of early-type galaxies are used as standard clocks. A sample of $ 31 $ data points for the redshift range $ 0.07<z<2.42 $ has been used for which the chi-square estimator has been found out to be
\begin{equation}
    \chi^2_{CC}=\sum_{i=1}^{31}\frac{[H_i^{th}(\theta_s,z_i)-H_i^{obs}(z_i)]^2}{\sigma^2_{CC}(z_i)}
\end{equation}
Here $ H_i^{th} $ is the theoretical value, with $ 
\theta_s $ being the vector of cosmological background parameters, $ H_i^{obs} $ is the observed value with $ \sigma_{CC} $ being the standard error in observed values. 

\subsection{Comparision with the $f(T, \mathcal{T})$ models}
\begin{widetext}

\begin{figure}[H]
\centering
    \captionsetup{width=.6\textwidth}
    \includegraphics[scale=0.6]{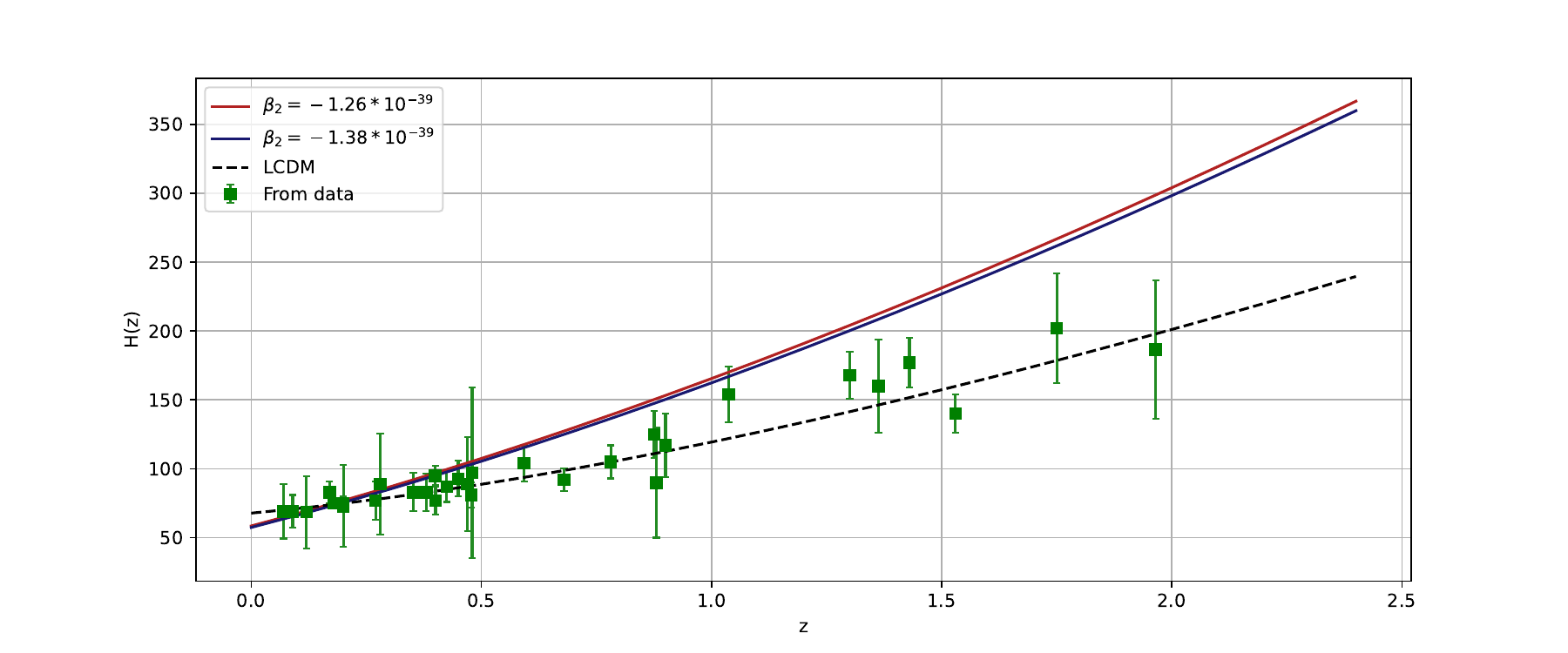}
    \caption{ The error bar plot of 31 points of Hubble datasets along with the Hubble model (for $f(T,\mathcal{T})= \beta_1 T+  \beta_2 \mathcal{T}$) compared with the $\Lambda CDM$ model. $\beta_2=-1.26 \times 10 ^{-39}$ represents the Red Curve and $\beta_2=-1.38 \times 10 ^{-39}$ represents the Blue Curve. }
     \label{fig 5}
\end{figure}
\end{widetext}

In Fig. \ref{fig 5}, we have plotted the Hubble model \eqref{35} for two different values of the model parameter $\beta_2$. Further, the curves have been compared with the $31$ points of the Hubble dataset and the standard $\Lambda$CDM model. The two model parameter values are the respective upper and lower bounds of the range of $\beta_2$, we obtained from BBN constraints. We observe that though this model fits well for the present time and a small range of redshift, for higher redshift it deviates from the observational datasets as well as the standard model. 

 \begin{widetext}

\begin{figure}[H]
   \centering
    \captionsetup{width=.6\textwidth}
    \includegraphics[scale=0.6]{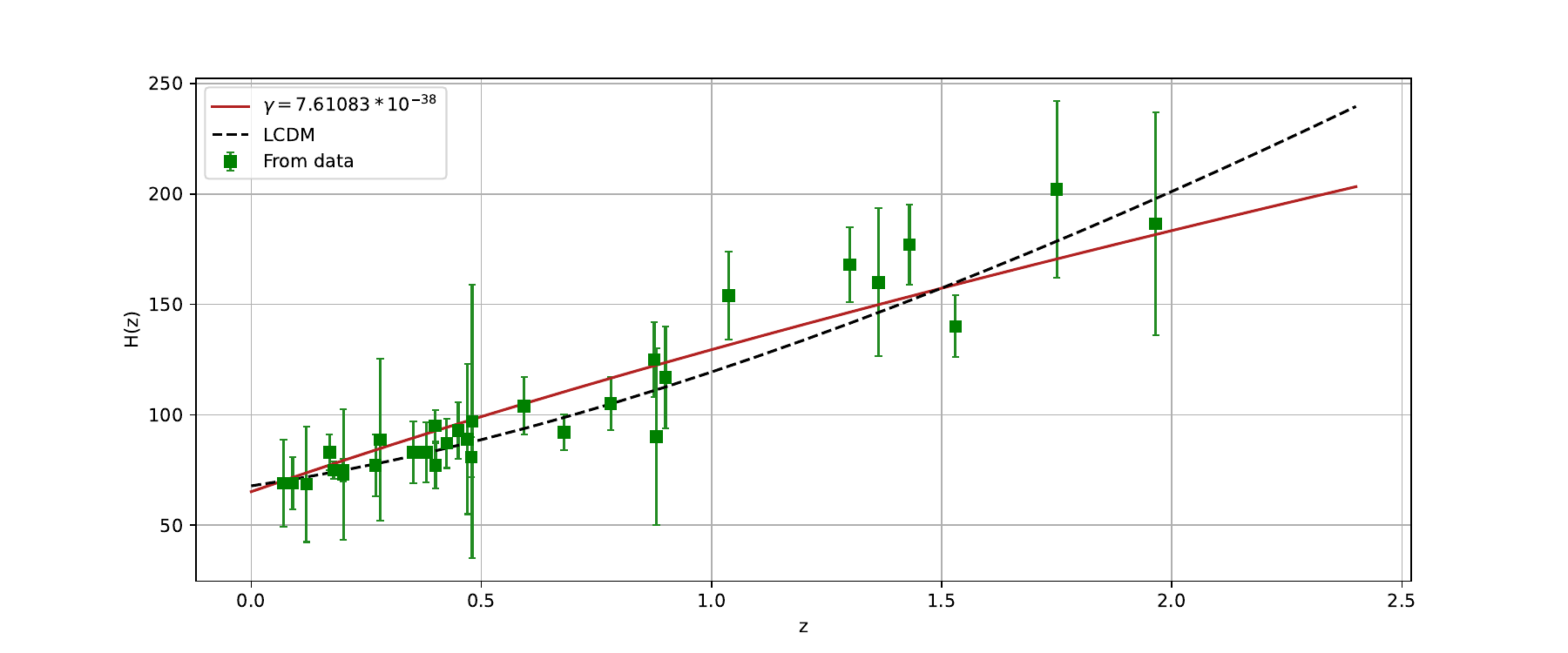}
 \caption{ The error bar plot of 31 points of Hubble datasets along with the Hubble model (for $f(T,\mathcal{T})=T+\alpha T^2 +\gamma \mathcal{T}$) compared with the $\Lambda CDM$ model.
 } 
   
     \label{fig 6}
\end{figure}
\end{widetext}

In Fig. \ref{fig 6}, we have plotted the Hubble model \eqref{41} for the model parameter $\gamma=7.61083\times10^{-38}$, which is obtained from the BBN constraints. Moreover, it has been compared with the $31$ points of the Hubble dataset and the standard $\Lambda$CDM model for the redshift range $0<z<2.5$. We infer from Fig. \ref{fig 6} that our model with the constrained model parameter, is compatible with the observational dataset and $\Lambda$CDM in describing the late time acceleration.\\

\section{Conclusion}
In this work, Big Bang Nucleosynthesis formalism and observations are employed to find the deviation in freeze-out temperature in the framework of $f(T, \mathcal{T})$ gravity. We assume a linear model and a non-linear one with squared torsion. In the first model, we find that to satisfy the constraint $|{\Delta T_f} / T_f|<4.7 \times 10^{-4}$, the corresponding model parameter ranges have to be $\beta_2 \in [-1.38\times 10^{-39},-1.26\times 10^{-39}]$ and $\beta_1 \in [-0.2640,-0.2615]$. 
Then the second model is explored using BBN constraints. We find that the model parameter $\gamma$ remains constant to satisfy BBN with the exact value $\gamma= 7.61083\times 10^{-38}$. Using this value we obtain the other model parameter $\alpha$ as $4.82\times 10^{82}$ Ge$V^{-2}$. 
\\
Further, we employ the constrained models from BBN epoch for a thorough comparison with the 31 data points of the Hubble dataset along with the standard $\Lambda$CDM model. We find that the linear model agrees with the observations for a small redshift range, whereas the nonlinear model  $(f(T,\mathcal{T})=T+\alpha T^2 +\gamma \mathcal{T})$ is an excellent candidate to explain the late time cosmic acceleration as well as the early BBN era. 

It should be noted that the increasing trend of the Hubble parameter in both the models with increasing redshift is to be expected since, even though the recession velocity of distant galaxies increases, the number of galaxies in a sphere of a fixed radius $ a(t) $ decreases with time causing the Hubble parameter to decrease with time or increase with redshift.

For the linear model, we observe that the contribution of the trace of the energy-momentum tensor corresponding to $ \beta_2 $ is 38 orders of magnitude less than that of torsion. This behaviour is expected since all the modified gravities must reproduce GR or the TEGR in this case, in a practical limit. As for the parameter $ \beta_1 \sim \mathcal{O}(1) $, we say that in order to reproduce the observed behaviour of the Universe in various epochs, the contribution from the torsion term has to be reduced by a certain amount. Following a similar line of reasoning, we can see that the parameter $ \alpha $ has an enormous value but since it is in units of $ GeV^{-2} $, its effective contribution to the action is very small. An even smaller contribution is that of the $ \mathcal{T} $ term with $ \gamma $ being $ \sim 10^{-38} $.

It is to be noted that the method of analysis employed here is a direct consequence of constraints obtained from the BBN era which are then imposed on the Hubble function corresponding to the specific $ f(T,\mathcal{T}) $ model. Since the BBN era exhibits accelerated expansion, it is expected that our constrained model also shows accelerated expansion for all values of redshift including the matter-dominated era which showed deceleration. The main reason behind this is the domination of dark energy components in the Hubble models. One might have to reduce our models to general relativity case to conduct a detailed analysis of the intermediating epochs that have shown decelerated expansion of the Universe.\\
We conclude that the model with a small correction to the linear model in the form of a squared torsion term can serve as a viable fit to explain both early time as well as late-time evolution of the Universe.\\

\textbf{Data availability} There are no new data associated with this article.

\acknowledgments  SSM acknowledges the Council of Scientific and Industrial Research (CSIR), Govt. of India for awarding Junior Research fellowship (E-Certificate No.: JUN21C05815).  PKS acknowledges Science and Engineering Research Board, Department of Science and Technology, Government of India for financial support to carry out Research project No.: CRG/2022/001847 and IUCAA, Pune, India for providing support through the visiting Associateship program. We are very much grateful to the honorable referee and to the editor for the illuminating suggestions that have significantly improved our work in terms
of research quality, and presentation.

\end{document}